# $Q$-РАСПРЕДЕЛЕНИЕ ДЛЯ ОДНОАТОМНОГО ЛАЗЕРА, РАБОТАЮЩЕГО В «КЛАССИЧЕСКОМ» РЕЖИМЕ

**Н.В. Ларионов**[1,2]

[1] Санкт-Петербургский государственный морской технический университет,
ул. Лоцманская, д. 3, Санкт-Петербург, 190121, Россия.
[2] Санкт-Петербургский политехнический университет Петра Великого,
ул. Политехническая, д. 29, Санкт-Петербург, 195251, Россия.

В работе теоретически исследуются модель одноатомного лазера с некогерентной накачкой. В стационарном случае из уравнения для оператора плотности системы выводится линейное однородное дифференциальное уравнение для усредненной по фазе $Q$-функции Хусими. В режиме когда связь поля с атомом во много раз сильнее, чем связь поля с резервуаром, обеспечивающим его распад, найдено асимптотическое решение этого уравнения. Это решение позволяет описать некоторые статистические особенности одноатомного лазера, в частности слабую суб-пуассоновскую статистику фотонов.

## Введение

В настоящее время источники неклассических состояний света востребованы в таких областях физики как квантовая информатика, квантовые коммуникации, квантовая криптография и квантовые стандарты частоты [1-5]. Проводятся различные исследования, направленные на создание таких источников. В частности, есть работы в которых, для получения определенных состояний света, а также для создания различных элементов квантовых устройств, предлагается использовать системы, состоящие всего из одного или нескольких квантовых излучателей [6-9]. Свойства одиночного излучателя отображаются на состоянии электромагнитного поля, что позволяет получить, к примеру, суб-пуассоновский свет [10].

Одной из фундаментальных моделей квантовой оптики является модель одноатомного лазера. Данной модели посвящено множество как теоретических [11-27], так и экспериментальных работ [28-30]. Различные эффекты, обнаруженные в этих работах, в основном связаны с сильным проявлением Ферми статистики одиночного излучателя: эффект самотушения, сжатие амплитудной компоненты поля, суб-пуассоновская статистика фотонов, генерация без инверсии и др.

Существенный вклад в понимание физики одноатомного лазера был сделан группой ученых из Института физики НАН Беларуси, возглавляемой профессором Килином С. Я. (см. [11], [15-18], [20,27] и ссылки там же). Один из теоретических подходов, используемый этой группой, основан на анализе уравнения для оператора плотности системы, записанного для таких квази-вероятностных распределений как $P$-функция Глаубера и $Q$-функция Хусими, позволяющих находить нормально- и антинормально упорядоченные корреляционные функции полевых операторов, соответственно.

В работе [21], для случая стационарного режима работы одноатомного лазера с некогерентной накачкой, было получено линейное однородное дифференциальное уравнение второго порядка для усредненной по фазе $P$-функции. В предельном случае, когда связь поля с атомом во много раз сильнее, чем связь поля с резервуаром, обеспечивающим его затухание (режим, в котором для одноатомного лазера возможно существование порога генерации [19], такого же, как и для обычного лазера - «классический» режим), было получено приближенное решение этого уравнение. Последнее, являющееся порождающим решением в проблеме малого параметра у старшей производной (сингулярно возмущенная задача, см., к примеру, [31]), для определенных значений параметров лазера дает хорошее согласие с численными расчетами и, более того, содержит в себе некоторые предельные решения, полученные ранее в [15,16]. Дальнейший анализ этого уравнения позволил получить приближенное выражение для $P$-функции, которая, в отличие от предыдущих решений, демонстрирует существенно неклассическое поведение - становится отрицательно определенной [22].

В силу специфики $P$-функции, которая может быть отрицательной и/или неограниченной, анализ упомянутого уравнения и его приближенных решений сталкивается с определенными трудностями. В частности здесь возникает проблема искусственного ограничения области определения $P$-функции (см., к примеру, [15], [21]). В связи с этим возникает естественное желание получить аналогичное уравнение, но для «хорошего» квази-вероятностного распределения. В качестве последнего, в представленной статье, было выбрано квази-вероятностное распределение $Q$, которое является неотрицательно определенным и ограниченным.

Структура статьи следующая. В первой части, для одноатомного лазера с некогерентной накачкой, генерирующего в стационарном режиме, выводится однородное дифференциальное уравнение для усредненной по фазе $Q$-функции. Вторая часть посвящена анализу этого уравнения в случае «классического» режима работы лазера. Находится асимптотическое решение этого уравнения, которое сравнивается с соответствующим решением для $P$-функции. С помощью найденного решения исследуется статистка фотонов в моде резонатора. Проводится сравнение с результатами, полученными в рамках подхода, основанного на линеаризации уравнений Гейзенберга-Ланжевена по малым флуктуациям вблизи сильного «классического» решения [19]. В заключении подводятся итоги проведенного исследования.

### Модель одноатомного лазера. Уравнение для усредненной по фазе *Q*-функции

Рассматриваемая модель одноатомного лазера представлена одиночным двухуровневым атомом, взаимодействующим с затухающей модой резонатора. Некогерентная накачка атома с нижнего уровня $|1\rangle$ на верхний уровень $|2\rangle$ осуществляется со скоростью $\Gamma/2$. Спонтанный распад атома с верхнего уровня $|2\rangle$ на нижний $|1\rangle$ происходит со скоростью $\gamma/2$. Константа взаимодействия атома с модой резонатора обозначается за $g$. Затухания моды резонатора происходит со скоростью $\kappa/2$.

Уравнение для оператора плотности $\hat{\rho}$ одноатомного лазера имеет вид

$$\begin{aligned}\frac{\partial\hat{\rho}}{\partial t} &= -\frac{i}{\hbar}\left[\hat{V},\hat{\rho}\right]+\frac{\kappa}{2}\left(2\hat{a}\hat{\rho}\hat{a}^{\dagger}-\hat{a}^{\dagger}\hat{a}\hat{\rho}-\hat{\rho}\hat{a}^{\dagger}\hat{a}\right)+\\ &+\frac{\gamma}{2}\left(2\hat{\sigma}\hat{\rho}\hat{\sigma}^{\dagger}-\hat{\sigma}^{\dagger}\hat{\sigma}\hat{\rho}-\hat{\rho}\hat{\sigma}^{\dagger}\hat{\sigma}\right)+\frac{\Gamma}{2}\left(2\hat{\sigma}^{\dagger}\hat{\rho}\hat{\sigma}-\hat{\sigma}\hat{\sigma}^{\dagger}\hat{\rho}-\hat{\rho}\hat{\sigma}\hat{\sigma}^{\dagger}\right),\end{aligned} \qquad (1)$$

где $\hat{a}^{\dagger},\hat{a}$ – операторы рождения и уничтожения фотонов в моде резонатора, соответственно; $\hat{\sigma}^{\dagger}=|2\rangle\langle 1|$, $\hat{\sigma}=|1\rangle\langle 2|$ – операторы атомных переходов; $\hat{V}=i\hbar g\left(\hat{a}^{\dagger}\hat{\sigma}-\hat{\sigma}^{\dagger}\hat{a}\right)$ – оператор взаимодействия атома с модой резонатора, где $\hbar$ – приведенная постоянная Планка. Физический смысл каждого слагаемого в правой части (1) определяется соответствующей скоростной константой.

Будем рассматривать антинормально упорядоченное диагональное представление оператора плотности $\hat{\rho}\left(z,z^{*}\right)$ по когерентным состояниям $|z\rangle$ поля, которое определяется следующим образом

$$\hat{\rho}=\int\hat{\rho}\left(z,z^{*}\right)|z\rangle\langle z|d^{2}z, \quad (2)$$

где $d^2 z \equiv d\,\mathrm{Re}[z]\,d\,\mathrm{Im}[z]$ и $\mathrm{Re}[z], \mathrm{Im}[z]$ - действительная и реальная часть комплексного числа $z$. Используя известные правила перехода [32]

$$\hat{a}\hat{\rho} \to \left( z + \frac{\partial}{\partial z^*} \right) \hat{\rho}\left(z, z^*\right),\ \hat{a}^\dagger \hat{\rho} \to z^* \hat{\rho}\left(z, z^*\right),$$
$$\hat{\rho}\hat{a} \to z\hat{\rho}\left(z, z^*\right),\ \hat{\rho}\hat{a}^\dagger \to \left( z^* + \frac{\partial}{\partial z} \right) \hat{\rho}\left(z, z^*\right), \qquad (3)$$

и вводя следующие функции: $\rho_{ik}\left(z, z^*\right) = \left\langle i \left| \hat{\rho}\left(z, z^*\right) \right| k \right\rangle,\ i, k = 1, 2$, $D\left(z, z^*\right) = \rho_{22}\left(z, z^*\right) - \rho_{11}\left(z, z^*\right)$ и $Q$-функцию $Q\left(z, z^*\right) = \rho_{11}\left(z, z^*\right) + \rho_{22}\left(z, z^*\right)$, из уравнения (1) можно легко получить систему дифференциальных уравнений в частных производных

$$\begin{cases} \dfrac{\partial Q}{\partial t} = \dfrac{\partial}{\partial z}\left[ \dfrac{\kappa}{2}\left( zQ + \dfrac{\partial Q}{\partial z^*} \right) - g\rho_{21} \right] + \dfrac{\partial}{\partial z^*}\left[ \dfrac{\kappa}{2}\left( z^* Q + \dfrac{\partial Q}{\partial z} \right) - g\rho_{12} \right], \\ \dfrac{\partial D}{\partial t} = (\Gamma - \gamma)Q - (\Gamma + \gamma)D + \dfrac{\partial}{\partial z}\left[ \dfrac{\kappa}{2} zD - g\rho_{21} \right] + \dfrac{\partial}{\partial z^*}\left[ \dfrac{\kappa}{2} z^* D - g\rho_{12} \right] - \\ -2g\left( z^* \rho_{21} + z\rho_{12} \right) + \kappa \dfrac{\partial^2 D}{\partial z \partial z^*}, \\ \dfrac{\partial \rho_{21}}{\partial t} = -\dfrac{(\Gamma + \gamma)}{2}\rho_{21} + \dfrac{\kappa}{2}\left[ \dfrac{\partial}{\partial z} z\rho_{21} + \dfrac{\partial}{\partial z^*} z^* \rho_{21} \right] + g\left[ zD + \dfrac{1}{2}\dfrac{\partial}{\partial z^*}(D - Q) \right] + \\ +\kappa \dfrac{\partial^2 \rho_{21}}{\partial z \partial z^*}. \end{cases} \qquad (4)$$

Здесь и далее, для того чтобы избежать громоздкости выражений, у квазивероятностей опущены скобки с комплексными аргументами $z, z^*$.

Физический смысл дополнительных функций $\rho_{21}$ и $D$ следует из их средних значений: $\langle D \rangle = \langle \rho_{22} \rangle - \langle \rho_{11} \rangle \equiv \langle \hat{\sigma}_z \rangle = \int D d^2 z$ – атомная инверсия, $\langle \rho_{21} \rangle = \langle \rho_{12} \rangle^* \equiv \langle \hat{\sigma} \rangle = \int \rho_{21} d^2 z$ – среднее значение атомной поляризации.

Первое уравнение в системе (4) можно переписать в виде уравнения неразрывности

$$\partial Q / \partial t + \mathrm{div}\,\mathbf{J} = q, \quad (5)$$

где определены дивергенция $\mathrm{div} = \left( \partial / \partial z, \partial / \partial z^* \right)$ и вектор тока квазивероятности $\mathbf{J} = \left( J, J^* \right)$, $J = -\kappa / 2\left( z + \partial / \partial z^* \right) Q$. Источник $q = -g\left( \partial \rho_{21} / \partial z + \partial \rho_{12} / \partial z^* \right)$ также может быть записан через дивергенцию некоторого вектора.

Перейдем к новым полярным координатам $(I,\varphi)$, таким что $z = I^{1/2}e^{i\varphi}$, и определим усредненные по фазе $\varphi$ квази-вероятностные распределения

$$\begin{gathered} Q(I) = \frac{1}{2\pi}\int_0^{2\pi} Q(I,\varphi)d\varphi,\ D(I) = \frac{1}{2\pi}\int_0^{2\pi} D(I,\varphi)d\varphi, \\ \rho_{12}(I) = \rho_{21}^*(I) = \frac{1}{2\pi}\int_0^{2\pi} e^{i\varphi}\rho_{12}(I,\varphi)d\varphi. \end{gathered} \qquad (6)$$

Тогда в стационарном случае из уравнения неразрывности (5) можно получить следующую связь между $Q(I)$ и суммой когерентностей $\rho_\Sigma(I) = \rho_{21}(I) + \rho_{12}(I)$

$$\rho_\Sigma(I) = \frac{\kappa}{g} I^{1/2}\left[Q(I) + \frac{dQ(I)}{dI}\right]. \quad (7)$$

Из двух последних уравнений системы (4), в стационарном случае, имеем

$$\begin{cases} (\Gamma-\gamma)Q(I) - (\Gamma+\gamma)D(I) - 2gI^{1/2}\rho_\Sigma(I) = \dfrac{d}{dI}\left[gI^{1/2}\rho_\Sigma(I) - \kappa I D(I) - \kappa I\dfrac{dD(I)}{dI}\right], \\ (\Gamma+\gamma)\rho_\Sigma(I) + \dfrac{\kappa}{2I}\rho_\Sigma(I) - 2gI^{1/2}\left[2D(I) + \dfrac{d}{dI}\big(D(I) - Q(I)\big)\right] = \\ = 2\kappa\dfrac{d}{dI}I\left[\rho_\Sigma(I) + \dfrac{d\rho_\Sigma(I)}{dI}\right]. \end{cases} \qquad (8)$$

Используя (7) можно исключить из системы (8) функцию $\rho_\Sigma(I)$ и получить систему двух дифференциальных уравнений относительно неизвестных функций $Q(I)$ и $D(I)$.

Наша задача состоит в получении одного дифференциального уравнения для функции $Q(I)$. Но первое уравнение в системе (8) (говорим о системе (8), подразумевая что $\rho_\Sigma(I)$ исключена из неё при помощи (7)) является дифференциальным уравнением второго порядка относительно функции $D(I)$, а второе - первого порядка относительно той же функции $D(I)$. Поэтому, для того чтобы исключить из (8) $D(I)$, $dD(I)/dI$ и $d^2D(I)/dI^2$, нужно добавить к этой системе еще одно уравнение, которое содержало бы вторую производную функции $D(I)$. Это уравнение можно получить, продифференцировав второе уравнение в системе (8).

Опуская промежуточные элементарные вычисления выпишем окончательный результат

$$\sum_{\nu=0}^{5} f_\nu(I) Q^{(\nu)}(I) = 0,$$
$$f_5(I) = b_{02}I^2 + b_{03}I^3,\ f_4(I) = b_{11}I + b_{12}I^2 + b_{13}I^3, \qquad (9)$$
$$f_3(I) = b_{20} + b_{21}I + b_{22}I^2 + b_{23}I^3,\ f_2(I) = b_{30} + b_{31}I + b_{32}I^2 + b_{33}I^3,$$
$$f_1(I) = b_{40} + b_{41}I + b_{42}I^2,\ f_0(I) = b_{50} + b_{51}I + b_{52}I^2,$$

где введено обозначение $Q^{(\nu)}(I) \equiv d^\nu Q(I)/dI^\nu$ и коэффициенты $b_{ik} = b_{ik}(\Gamma, \gamma, \kappa, g)$ выписаны в приложении.

Таким образом, искомое уравнение для усредненной по фазе *Q*-функции является однородным дифференциальным уравнением пятого порядка с полиномиальными коэффициентами. Напомним, что соответствующее уравнение для усредненной по фазе *P*-функции, является уравнением второго порядка [21].

**Функция *Q*(*I*) для одноатомного лазера, работающего в «классическом» режиме**

Далее, как и в работах [19,21], будем использовать следующие три безразмерных параметра: безразмерный параметр накачки - $r = \Gamma / \gamma$, безразмерный коэффициент насыщения - $I_s = \gamma / \kappa$ и безразмерную константу связи (кооперативный параметр) - $c = 4g^2 / \gamma\kappa$ .

Как упоминалось выше, для рассматриваемого одноатомного лазера было выведено уравнение для усредненной по фазе *P*-функции $P(I)$ [21]. Это уравнение представляет собой однородное дифференциальное уравнение второго порядка с полиномиальными коэффициентами. В «классическом» режиме, когда произведение $cI_s >> 1$ (т.е. $g / \kappa >> 1$), в этом уравнении можно выделить малый параметр $\lambda \approx 1/ cI_s$, стоящий при старшей производной. Используя теорию возмущений, авторы статьи нашли порождающее решение $P_0(I)$ этого уравнения (формула (50) в [21]), являющееся решением дифференциального уравнения первого порядка.

В случае «хорошего» резонатора $I_s >> 1$ функция $P_0(I)$ хорошо описывает статистические свойства одноатомного лазера. В частности, с её помощью можно описать некоторые результаты, получающиеся из решения системы уравнений Гейзенберга-Ланжевена путем их линеаризации по малым флуктуациям вблизи сильного «классического» решения [19,21] (далее «линейная теория»). Выпишем результаты этой «линейной теории»

$$I_0\left(r,I_s,c\right)=\frac{I_s}{2}\left[\left(r-1\right)-\frac{\left(r+1\right)^2}{c}\right],$$
$$Q_f^{lin}\left(r,c\right)=\frac{2c^2-c(r-5)(r+1)+3(r+1)^3}{2c^2\left[(r-1)-\frac{(r+1)^2}{c}\right]}. \quad (10)$$

Здесь $I_0\approx\langle\hat{n}\rangle\equiv\langle\hat{a}^\dagger\hat{a}\rangle$ - классическая внутрирезонаторная интенсивности, а $Q_f^{lin}\approx Q_f=\left(\langle\hat{n}^2\rangle-\langle\hat{n}\rangle^2\right)/\langle\hat{n}\rangle-1$ - $Q$-параметр Манделя для поля (верхний индекс «*lin*» указывает на то, что это результат «линейной теории»).

Формулы (10), имеют смысл для $c>8$ и для $r\in\left(r_{th},r_q\right)$, где $r_{th}$ - пороговое значение параметра накачки и $r_q$ - значение параметра накачки, при котором происходит эффект самотушения. Явные выражения для $r_{th},r_q$ получаются из уравнения $I_0=0$ и они равны

$$r_{th}=r_m-\frac{c}{2}\sqrt{1-\frac{8}{c}},\ r_q=r_m+\frac{c}{2}\sqrt{1-\frac{8}{c}}, \quad (11)$$

где $r_m=c/2-1$ значение накачки, когда интенсивность $I_0$ достигает своего максимума $I_m=I_s\left(c/8-1\right)$.

Результаты (10) предсказывают незначительную суб-пуассоновскую статистику фотонов в моде резонатора [21]. При $c\approx 200$ и выше, и для значений параметра накачки близких к $r=c/5$, Мандель $Q$-параметр $Q_f^{lin}$ становится отрицательным и при $c\to\infty$ принимает значение равное $-0.05$ (проявление эффекта антигруппировки фотонов). $P$-функция в этой области значений параметров демонстрирует неклассическое поведение - становится отрицательно определенной и неограниченной, а у приближенного решения $P_0\left(I\right)$ появляется существенно особая точка близкая к $I_0$, не позволяющая подтвердить результаты «линейной теории».

Отметим, что для рассматриваемой модели одноатомного лазера выявленное минимальное значение $Q$-параметра Манделя равно $-0.15$ [13], [24]. Такая суб-пуассоновская статистика связана с эффектом антигруппировки фотонов, который лучше всего проявляется в режиме малого числа фотонов в моде резонатора $rI_s\approx 1$ ($\Gamma\approx\kappa$) [24]. Однако в рассматриваемом нами «классическом» режиме, т.е. в режиме когда существуют решения (10), этот эффект ослаблен присутствием большого числа фотонов, некогерентно накопленных в моде и обладающих относительно длительным временем жизни.

Теперь перейдем к выделению малого параметра $\lambda \approx 1/cI_s$ в уравнении (9) и попробуем получить соответствующее порождающее решение для *Q*-функции. Будем действовать также как и в работе [21]. Для этого перепишем полиномы $f_\nu(I)$ выделив в них корни

$$
\begin{aligned}
&f_5(I) = b_{03}I^2(I - I_{00}),\ f_4(I) = b_{13}I(I - I_{11})(I - I_{12}),\\
&f_3(I) = b_{23}(I - I_{21})(I - I_{22})(I - I_{23}),\ f_2(I) = b_{33}(I - I_{31})(I - I_{32})(I - I_{33}),\\
&f_1(I) = b_{42}(I - I_{-4})(I - I_{+4}),\ f_0(I) = b_{52}(I - I_{-5})(I - I_{+5}),
\end{aligned}
\qquad (12)
$$

где в последних двух полиномах обозначения корней аналогичны обозначениям в работе [21].

Коэффициенты при двух последних полиномах $b_{42}, b_{52} \sim cI_s$, а коэффициенты при всех остальных полиномах порядка единицы. Поэтому в режиме $cI_s >> 1$ заменим уравнение (9) на приближенное уравнение первого порядка

$$b_{42}(I - I_{-4})(I - I_{+4})Q_0'(I) + b_{52}(I - I_{-5})(I - I_{+5})Q_0(I) = 0. \quad (13)$$

Решение этого уравнения имеет следующий вид

$$
\begin{aligned}
&Q_0(I) = N_0(1 - I/I_{-4})^{f_1}\left|(1 - I/I_{+4})\right|^{f_2}\exp\left(-\frac{b_{52}}{b_{42}}I\right),\\
&f_1 = -\frac{b_{52}}{b_{42}}\frac{(I_{-4} - I_{-5})(I_{-4} - I_{+5})}{(I_{-4} - I_{+4})},\ f_2 = \frac{b_{52}}{b_{42}}\frac{(I_{+4} - I_{-5})(I_{+4} - I_{+5})}{(I_{-4} - I_{+4})},
\end{aligned}
\qquad (14)
$$

где $N_0$ - нормировочная константа, а нижний индекс «*0*» у функции указывает на то, что это решение является порождающим в задаче с малым параметром.

Найденное решение (14) по своей структуре практически полностью совпадает с соответствующим порождающим решением $P_0(I)$ для *P*-функции (см. формула (50) в [21]). Однако у решения (14) есть некоторые преимущества, связанные с тем, что *Q*-функция неотрицательно определена и ограниченна. Так экспонента в (14), в связи с положительностью отношения $b_{52}/b_{42} > 0$, убывает при возрастании аргумента *I*. Напротив, функция $P_0(I)$, найденная в [21], экспоненциально возрастает при $I \to \infty$, что являлось одной из причин искусственного ограничения ее области определения.

Первая скобка $\left(1-I/I_{-4}\right)$ в (14) всегда положительна, так как корень $I_{-4}<0$. Вторая скобка $\left(1-I/I_{+4}\right)$ становится отрицательной для $I>I_{+4}$ (корень $I_{+4}>0$), а, следовательно, $Q_0\left(I\right)$ принимает комплексное значение, что недопустимо для $Q$-функции. Однако для случая «хорошего» резонатора $I_s>>1$ значение корня $I_{+4}$ близко к тем значениям переменной $I$, для которых $Q$-функция пренебрежимо малая величина. Для случая «плохого» резонатора $I_s\sim 1, I_s<<1$ значение корня $I_{+4}$ находится в области значений переменной $I$, где $Q$-функция не является малой. Если же ограничивать область определения функции $Q_0\left(I\right)$ отрезком $\left[0,I_{+4}\right]$, как это было сделано для функции $P_0\left(I\right)$ в [21], то это может приводить к нефизическим результатам. Поэтому скобка $\left(1-I/I_{+4}\right)$ в (14) взята под знак модуля.

Отметим несколько особенностей решения (14), которые есть и у функции $P_0\left(I\right)$. Корень $I_{-5}$ практически полностью повторяет решение для классической внутрирезонаторной интенсивности (10), т.е. $I_{-5}\approx I_0$. В случае «хорошего» резонатора, вблизи максимума классического решения $I_0$, выполняется приближенное равенство $I_{+4}\approx I_{+5}$.

Анализ уравнения (9) показывает, что решение (14) не может быть использовано для описания работы лазера в случае, когда параметр накачки много меньше порогового значения $r_{th}$. В этом случае, учитывая что основные изменения $Q$-функции происходят в области малых значений переменной $I$, можно рассмотреть уравнение, получающееся из (9) путем пренебрежения в полиномах всеми степенями переменной $I$, т.е.

$$b_{20}Q^{(3)}\left(I\right)+b_{30}Q^{(2)}\left(I\right)+b_{40}Q^{(1)}\left(I\right)+b_{50}Q\left(I\right)=0. \quad (15)$$

Решение этого уравнения

$$Q_1\left(I\right)=N_0e^{I/a}, \quad (16)$$

где константа $a<0$ является действительным корнем полиномиального уравнения: $b_{50}+b_{40}I+b_{30}I^2+b_{20}I^3=0$. Эта константа при $r\to 0,\infty$ стремиться к минус единице, т.е. в этих предельных случаях $Q$-функция описывает вакуумное состояния моды - $Q_1\left(I\right)=e^{-I}/\pi.$ Таким образом, решение (16) описывает тепловое излучение со средним числом фотонов равным $\left\langle\hat{n}\right\rangle=-\left(a+1\right)$.

## Результаты расчетов

Ни рисунке 1 сравниваются поведения функций $P_0(I)$ из [21] и $Q_0(I)$ (14) при переходе в режим близкий к режиму сильной связи $c >> I_s$ (т.е. $g >> \gamma$ - связь атома с полем во много раз сильнее, чем связь атома с резервуаром, обеспечивающим его спонтанный распад вне моду резонатора). Для всех графиков $r = c/5$, т.е. выбрано значение параметра накачки при котором Мандель $Q$-параметр $Q_f^{lin}$ (10) принимает своё минимальное значение. Видно, что увеличение константы связи $c$, при фиксированном $I_s$, приводит к всё более узкому и выраженному пику для функции $P_0(I)$. При $c = 175$ заметна несимметричность этой функции, связанная с ограниченностью ее области определения. Для $c \approx 200$ у функции $P_0(I)$ появляется особенность (аналогичная особенность наблюдалась в [24,25]), которая не позволяет вычислить интересующие средние величины. При дальнейшем увеличении константы связи $c$ эта особенность уходит за область определения функции $P_0(I)$, но вычисление средних дает сомнительные результаты. Функция $Q_0(I)$, как и следовало ожидать, не имеет никаких особенностей и позволяет легко вычислять интересующие средние величины.

Для значений параметров лазера, соответствующих рисункам 1 a),b),c), вычисленные с помощью функций $P_0(I)$, $Q_0(I)$ средние значения фотонов и Мандель $Q$-параметры с хорошей точностью совпадают с результатами «линейной» теории (10) (*H-L*):

$$
\begin{array}{llll}
 & a)\ I_s = 100,\ c = 50 & b)\ I_s = 100,\ c = 100 & c)\ I_s = 100,\ c = 175 \\
H\text{-}L & I_0 = 329,\ Q_f^{lin} = 0.19 & I_0 = 729.5,\ Q_f^{lin} = 0.056 & I_0 = 1329.7,\ Q_f^{lin} = 0.0075 \\
P_0(I) & \langle n \rangle = 329.1,\ Q_f = 0.19 & \langle n \rangle = 729.6,\ Q_f = 0.056 & \langle n \rangle = 1329.8,\ Q_f = 0.008 \\
Q_0(I) & \langle n \rangle = 329.1,\ Q_f = 0.19 & \langle n \rangle = 729.6,\ Q_f = 0.057 & \langle n \rangle = 1329.8,\ Q_f = 0.0084
\end{array}.
$$

На рисунке 1 d), все также для $r = c/5$, представлено поведение функции $Q_0(I)$ при переходе в режим сильной связи. При этом параметры $c$, $I_s$ подобраны так, чтобы классическая внутрирезонаторная интенсивность была равна $I_0 = 700$.

Выпишем соответствующие значения $I_0, Q_f^{lin}$ и $\langle n\rangle, Q_f$:

| | $I_s = 95.95,\ c = 10^2$ | $I_s = 8.83,\ c = 10^3$ | $I_s = 0.87,\ c = 10^4$ |
|---|---|---|---|
| $H\text{-}L$ | $I_0 = 700,\ Q_f^{lin} = 0.056$ | $I_0 = 700,\ Q_f^{lin} = -0.04$ | $I_0 = 700,\ Q_f^{lin} = -0.049$ |
| $P_0(I)$ | $\langle n\rangle = 700.1,\ Q_f = 0.056$ | $\langle n\rangle = —,\ Q_f = --$ | $\langle n\rangle = —,\ Q_f = --$ |
| $Q_0(I)$ | $\langle n\rangle = 700.1,\ Q_f = 0.057$ | $\langle n\rangle = 700.1,\ Q_f = -0.039$ | $\langle n\rangle = 700.1,\ Q_f = -0.047$ |

В этой таблице только первые два столбца соответствуют графикам на рисунке 1 d). Последний столбец относится к режиму сильной связи и «плохого» резонатора (график не приведён, так как он практически полностью совпадает с графиком для случая $c = 10^3$). Из приведенных значений видно, что найденная $Q$-функция (14) хорошо описывает слабую суб-пуассоновскую статистику, предсказанную «линейной» теорией. Случай «плохого» резонатора и случаи $c \approx 200,\ c > 200$ функция $P_0(I)$ описать не может.

На рисунке 2 сравниваются результаты «линейной» теории с результатами, полученными с помощью решений (14), (16). Для случая «хорошего» резонатора $I_s >> 1$ и для значений параметра накачки, лежащих между $r_{th}$ и $r_q$ (рисунок 2 a), b), c)), оба подхода полностью согласуются друг с другом. Характерный пороговый пик для $Q$-параметра Манделя [19] и пик, обусловленный эффектом «запирания» атома в возбужденном состоянии, хорошо описывается найденной функцией $Q_0(I)$ (14). Для значений параметра накачки лежащих много ниже классического порога $r_{th}$, решение (14) даёт нефизический результат для $Q$-параметра Манделя. Для получения результатов в этом случае использовано решение (16) (синяя точечная кривая).

На рисунке 2 c) рассматривается переход в режим сильной связи, когда возникает слабая суб-пуассоновская статистика фотонов, предсказанная «линейной» теорией. Как видно из рисунка, найденная функция $Q_0(I)$ (14) полностью описывает этот квантовый эффект.

Режим «плохого» резонатора $I_s \sim 1$ рассмотрен на рисунке 2 d). Графики для $Q$-параметра Манделя $Q_f^{lin}$ (10), в силу его естественной зависимости только от параметров $r, c$, полностью совпадают с графиками на рисунке 2c). Из сравнения с численными расчетами видно, что результаты, полученные с помощью «линейной» теории (10), лучше описывают переход в суб-пуассоновский режим. Для «плохого» резонатора функция $Q_0(I)$ (14) дает только качественный результат. Однако в пределе $c \to \infty$ результаты полученные с помощью функции $Q_0(I)$ (14) совпадут с результатами «линейной» теории.

**Аппроксимация функции $Q_0(I)$ гауссовым распределением**

Рассмотрим режим «хорошего» резонатора $I_s >> 1$ и значения параметра накачки близки к $r_m$ (10), (11). Учтем, что в рассматриваемом случае $I_{+4} \approx I_{+5}$ и $I_{-5} \approx I_0$. Отсюда $f_1 \approx -b_{52} / b_{42} (I_{-4} - I_0)$, $f_2 \approx 0$ и (14) можно переписать следующим образом

$$Q_0(I) = N_0 \left(1 + \frac{\Delta I}{I_0 - I_{-4}}\right)^{\frac{b_{52}}{b_{42}}(I_0 - I_{-4})} \exp\left(-\frac{b_{52}}{b_{42}} \Delta I\right), \quad (17)$$

где $\Delta I = (I - I_0)$ и величина $(I_0 - I_{-4}) >> 1$. (17) имеет максимум соответствующий значениям $I$, близким к классическому решению $I_0$, т.е. для $\Delta I \approx 0$. Тогда в области значений переменной $I$, для которых (17) не является пренебрежимо малой, т.е. для $I$ не слишком удаленных от значения $I_0$, можно сделать следующее разложение предэкспоненциального множителя

$$\begin{aligned} \left(1 + \frac{\Delta I}{I_0 - I_{-4}}\right)^{-(I_0 - I_{-4})} &= \exp\left[-(I_0 - I_{-4}) \ln\left(1 + \frac{\Delta I}{I_0 - I_{-4}}\right)\right] \approx \\ &\approx \exp\left[-\Delta I + \frac{(\Delta I)^2}{2(I_0 - I_{-4})}\right]. \end{aligned} \quad (18)$$

Подставляя (18) в (17) видим что $Q$-функция принимает вид гауссовой функции

$$Q_0(I) = N_0 \exp\left[-\frac{(\Delta I)^2}{2\sigma^2}\right]. \quad (19)$$

где дисперсия $\sigma^2 = (I_0 - I_{-4}) b_{42} / b_{52}$ определяется решениями (10)

$$\sigma^2 = \pi \int_0^\infty I^2 Q_0(I) dI - \left[\pi \int_0^\infty I Q_0(I) dI\right]^2 = 1 + I_0 \left(2 + Q_f^{lin}\right). \quad (20)$$

Таким образом, как это следует из (20), решение (19) описывает слабую суб-пуассоновскую статистику фотонов в моде резонатора. В самом деле, по определению Манделя $Q$-параметра, имеем

$$Q_f = \frac{\pi\int\limits_0^\infty I^2 Q(I)\,dI - \left[\pi\int\limits_0^\infty IQ(I)\,dI\right]^2 - \pi\int\limits_0^\infty IQ(I)\,dI}{\pi\int\limits_0^\infty IQ(I)\,dI - 1} - 1 \approx$$

$$\approx \frac{\sigma^2 - I_0 - 1}{I_0} - 1 = Q_f^{lin}. \quad (21)$$

Отметим, что для функции $P_0(I)$ в работе [21] также было получено гауссово выражение, типа (19). Однако дисперсия $\sigma^2$ в этом случае была равна следующему произведению $I_0 \cdot Q_f^{lin}$, которое для суб-пуассоновской статистики становилось отрицательным, что делало функцию $P_0(I)$ неограниченной.

## Заключение

В данной работе, на основе уравнения для усредненной по фазе *Q*-функции (9), исследовался стационарный режим работы одноатомного лазера с некогерентной накачкой. В «классическом» режиме ($g/\kappa >> 1$) получено приближенное решение этого уравнения (14), (16). Это решение описывает основные особенности одноатомного лазера, в частности позволяет описать слабую суб-пуассоновскую статистику фотонов в моде резонатора, ранее обнаруженную с помощью подхода, основанного на линеаризации уравнений Гейзенберга – Ланжевена вблизи сильного классического решения (10) [19,21].

Подытоживая можно сказать, что анализ стационарного режима работы одноатомного лазера с некогерентной накачкой может быть сведен к анализу одного линейного однородного дифференциального уравнения. Так в случае *P*-функции это уравнение (41) в [21], а для *Q*-функции – уравнение (9) данной статьи. В рассмотренном здесь и в [21] предельном случае $g/\kappa >> 1$ в упомянутых дифференциальных уравнениях появляется малый параметр, который относительно легко позволяет найти приближенные решения этих уравнений.

В конце отметим, что уравнение (9) также анализировалось в работе [25], где рассматривался частный случай $\Gamma \approx \kappa$, т.е. случай малого числа фотонов в моде резонатора. Для произвольных значений отношения $g/\kappa$ анализ уравнения (9) был затруднителен. Только в предельном случае $g/\Gamma \approx g/\kappa \to \infty$ были найдены точные аналитические решения уравнения (9), одно из которых совпадает с решением (16). В случае малого числа фотонов более плодотворным оказался подход, основанный на анализе бесконечной системы алгебраических уравнений для различных моментов полевых операторов [33]. Возможно именно этот подход, для некоторых частных случаях, позволит найти точные решения для средних величин, характеризующих одноатомный лазер [34].

Работа выполнена в рамках Государственного задания на проведение фундаментальных исследований (код темы FSEG-2020-0024).

## Приложение

Коэффициенты $b_{ik} = b_{ik}\left(\omega,\eta,\tau\right)$ из уравнения (9), где для упрощения введены следующие безразмерные константы $\omega = \Gamma / 2g,\ \eta = \gamma / 2g,\ \tau = \kappa / 2g$

$$b_{02} = -2\tau^3\left(\omega+\eta+\tau\right), b_{03} = 4\tau^4;$$
$$b_{11} = -12\tau^3\left(\omega+\eta+\tau\right), b_{12} = 2\tau^3\left(7\tau-3\omega-3\eta\right), b_{13} = 12\tau^4;$$
$$b_{20} = -12\tau^3\left(\omega+\eta+\tau\right), b_{21} = -\tau^2\left(26\eta\tau-3\eta^2+21\tau^2-6\eta\omega+26\tau\omega-3\omega^2\right),$$
$$b_{22} = 12\tau^3\left(4\tau-\omega-\eta\right), b_{23} = 12\tau^4;$$
$$b_{30} = -2\tau^2\left(8\eta\tau-3\eta^2+15\tau^2-6\eta\omega+8\tau\omega-3\omega^2\right);$$
$$b_{31} = -2\tau(\eta+\tau-3\eta^2\tau+13\eta\tau^2+\omega-6\eta\tau\omega+13\tau^2\omega-3\tau\omega^2),$$
$$b_{32} = 2\tau^2\left(2-7\eta\tau+23\tau^2-7\tau\omega\right), b_{33} = 4\tau^4;$$
$$b_{40} = -\eta^3\tau+\eta^2(8\tau^2-1-3\tau\omega)-\tau(3\tau+24\tau^3+3\omega-\tau^2\omega-8\tau\omega^2+\omega^3)+$$
$$+\eta(\tau^3-\omega+16\tau^2\omega-\tau(4+3\omega^2)),$$
$$b_{41} = \tau(5\eta^2\tau+15\tau^3-4\omega-20\tau^2\omega-2\eta(1+10\tau^2-5\tau\omega)+\tau(5\omega^2-2)),$$
$$b_{42} = 2\tau^2(4-3\eta\tau+7\tau^2-3\tau\omega);$$
$$b_{50} = -\eta^3\tau-6\tau^4+5\tau^3\omega+\omega^2-\tau\omega^3+\eta^2(2\tau^2-1-3\tau\omega)+\eta\tau(5\tau^2-4+4\tau\omega-3\omega^2)+\tau^2(2\omega^2-3),$$
$$b_{51} = 2\tau(\eta^2\tau+3\tau^3-2\omega-4\tau^2\omega+\tau\omega^2+2\eta\tau(\omega-2\tau)), b_{52} = 4\tau^2.$$

**СПИСОК ЛИТЕРАТУРЫ**

1. I. R. Berchera and I. P. Degiovanni, Metrologia **56,** 024001 (2019).

2. V. D’ambrosio, N. Spagnolo, L. Del Re, et al., Nat. Commun. **4**, 2432 (2013).

3. S. Pogorzalek, K. G. Fedorov, M. Xu, et al., Nat. Commun. **10**, 2604 (2019).

4. К. С. Тихонов, А. Д. Манухова, С. Б. Королёв, Т. Ю. Голубева, Ю. М. Голубев, Опт. и спектр. **127** (11), 811 (2019).

5. J. Shi, G. Patera, D. B. Horoshko and M. I. Kolobov, J. Opt. Soc. Am. B **37,** 3741 (2020).

6. S. Ritter, C. Nölleke, C. Hahn, et al., Nature **484**, 195 (2012).

7. С. О. Тарасов, С. Н. Андрианов, Н. М. Арсланов, С. А. Моисеев, Известия РАН. Серия физическая **82** (8), 1148 (2018).

8. Е. Н. Попов, В. А. Решетов, Письма в ЖЭТФ **111**, 846 (2020).

9. A. A. Sokolova, G. P. Fedorov, E. V. Il'ichev and O. V. Astafiev, Phys. Rev. A **103**, 013718 (2021).

10. Д. Ф. Смирнов, А. С. Трошин, УФН **153**, 233 (1987).

11. S. Ya. Kilin and T. B. Krinitskaya, J. Opt. Soc. Am. B **8**, 2289 (1991).

12. Yi Mu and C. M. Savage, Phys. Rev. A **46**, 5944 (1992).

13. А. В. Козловский, А. Н. Ораевский, ЖЭТФ **115**, 1210 (1999).

14. B. Jones, S. Ghose, J. P. Clemens, P. R. Rice, and L. M. Pedrotti, Phys. Rev. A **60**, 3267 (1999).

15. Т. Б. Карлович, С. Я. Килин, Опт. и спектр. **91**, 374 (2001).

16. С. Я. Килин, Т. Б. Карлович, ЖЭТФ **122**, 933 (2002).

17. Т. Б. Карлович, С. Я. Килин, Опт. и спектр. **103**, 288 (2007).

18. Т. Б. Карлович, Опт. и спектр. **111**, 758 (2011).

19. N. V. Larionov and M. I. Kolobov, Phys. Rev. A **84**, 055801 (2011).

20. S. Ya. Kilin and A. B. Mikhalychev, Phys. Rev. A **85**, 063817 (2012).

21. N. V. Larionov and M. I. Kolobov, Phys. Rev. A **88**, 013843 (2013).

22. E. N. Popov and N. V. Larionov, Proc. SPIE **9917**, 99172X (2016). DOI: 10.1117/12.2229228

23. В. А. Бобрикова, Р. А. Хачатрян, К. А. Баранцев, Е. Н. Попов, Опт. и спектр. **127**, 976 (2019).

24. N. V. Larionov, Proc. 2020 IEEE International Conference on Electrical Engineering and Photonics (EExPolytech), 265 (2020). DOI: 10.1109/EExPolytech50912.2020.9243955

25. N. V. Larionov, Journal of Physics: Conference Series, **2103**, 012158 (2021), DOI: 10.1088/1742-6596/2103/1/012158

26. B. Parvin, Eur. Phys. J. Plus **136**, 728 (2021).

27. Dmitri B. Horoshko, Chang-Shui Yu, and Sergei Ya. Kilin, J. Opt. Soc. Am. B **38**, 3088 (2021).

28. J. McKeever, A. Boca, A. D. Boozer, J. R. Buck, and H. J. Kimble, Nature **425**, 268 (2003).

29. M. Nomura, N. Kumagai, S. Iwamoto, Y. Ota, and Y. Arakawa, Opt. Express **17**, 15975 (2009).

30. F. Dubin, C. Russo, H. G. Barros A. Stute, C. Becher, P. O. Schmidt, and R. Blatt, Nat. Phys. **6**, 350 (2010).

31. Найфэ, А. Введение в методы возмущений, Москва : Мир, 1984.

32. Мандель, Л. Оптическая когерентность и квантовая оптика : пер. с англ. / Л. Мандель, Э. Вольф. Москва : Физматлит, 2000.

33. G. S. Agarwal and S. Dutta Gupta, Phys. Rev. A **42**, 1737 (1990).

34. Ф. М. Федоров, Бесконечные системы линейных алгебраических уравнений и их приложения, Наука, Новосибирск (2011).

## ПОДПИСИ К РИСУНКАМ

Рис. 1. Сравнение поведений функций $P_0(I)$ [21] и $Q_0(I)$ (14): a) $I_s = 100,\ c = 50$; b) $I_s = 100,\ c = 100$; c) $I_s = 100,\ c = 175$. d) - поведение функции $Q_0(I)$ (14).

Рис. 2. Сравнение результатов «линейной» теории (10) с результатами, полученными с помощью функции $Q_0(I)$ (14). a) и b) - зависимости среднего числа фотонов $\langle n(r) \rangle$ и Манделя Q-параметра $Q_f(r)$ от параметра накачки $r$. $I_s = 40,\ c = 20$. c) и d) - зависимость Манделя Q-параметра $Q_f(r)$ от параметра накачки $r$. Переход в режим сильной связи: c) - случай «хорошего» резонатора, $I_s = 40$; d) - случай «плохого» резонатора, $I_s = 1$. Зеленные точки – численные счет уравнения (1).

## РИСУНКИ

**Рис. 1:**

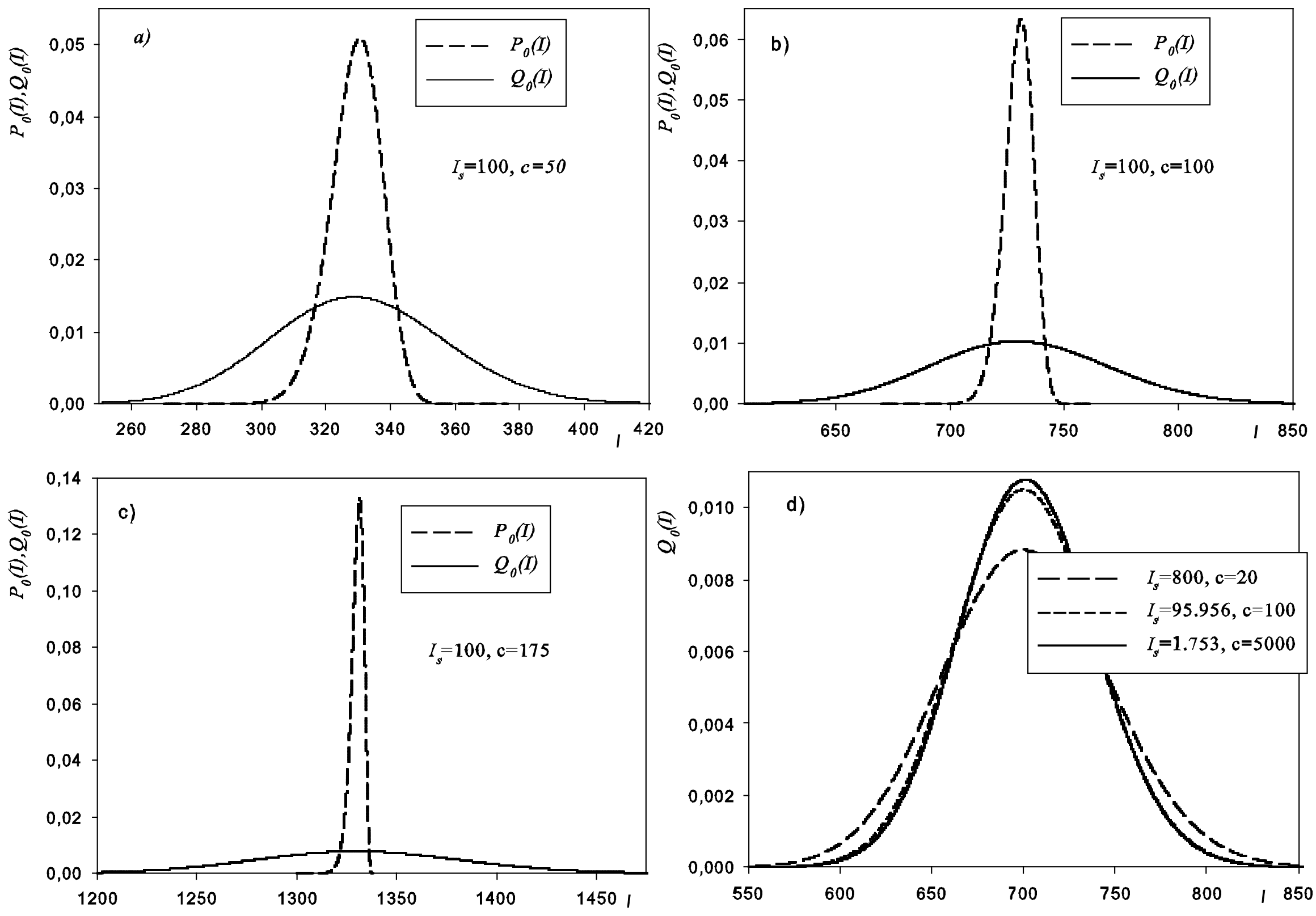

**Рис. 2:**

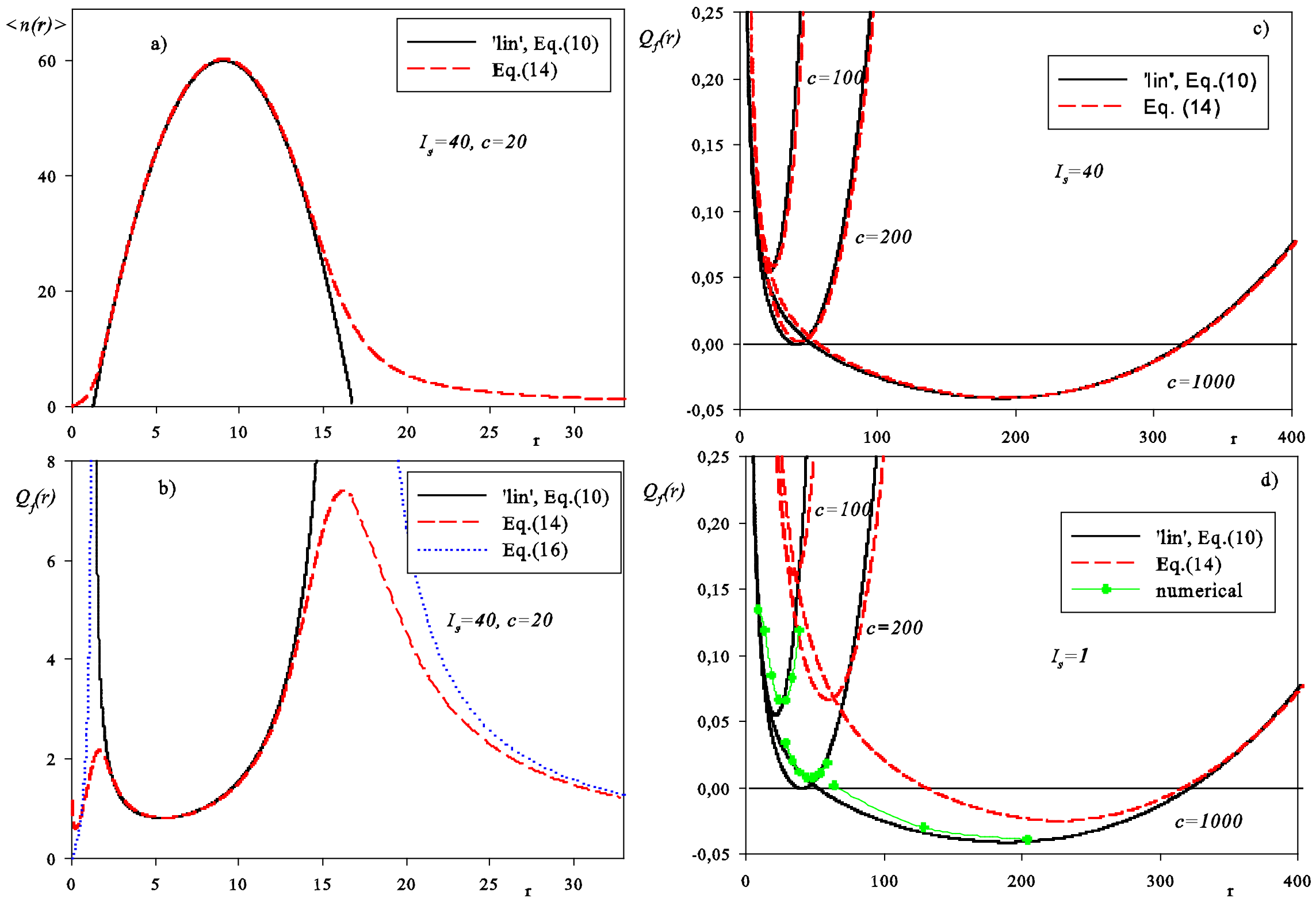